# On Box-Cox Transformation for Image Normality and Pattern Classification

ABBAS CHEDDAD  (Senior Member, IEEE)

Department of Computer Science, Blekinge Institute of Technology, SE-371 79 Karlskrona, Sweden

e-mail: abbas.cheddad@bth.se

This work was supported by the Swedish Knowledge Foundation, scalable resource efficient systems for big data analytics, under Grant 20140032.

**ABSTRACT** A unique member of the power transformation family is known as the Box-Cox transformation. The latter can be seen as a mathematical operation that leads to finding the optimum lambda ($\lambda$) value that maximizes the log-likelihood function to transform a data to a normal distribution and to reduce heteroscedasticity. In data analytics, a normality assumption underlies a variety of statistical test models. This technique, however, is best known in statistical analysis to handle one-dimensional data. Herein, this paper revolves around the utility of such a tool as a pre-processing step to transform two-dimensional data, namely, digital images and to study its effect. Moreover, to reduce time complexity, it suffices to estimate the parameter lambda in real-time for large two-dimensional matrices by merely considering their probability density function as a statistical inference of the underlying data distribution. We compare the effect of this light-weight Box-Cox transformation with well-established state-of-the-art low light image enhancement techniques. We also demonstrate the effectiveness of our approach through several test-bed data sets for generic improvement of visual appearance of images and for ameliorating the performance of a colour pattern classification algorithm as an example application. Results with and without the proposed approach, are compared using the AlexNet (transfer deep learning) pretrained model. To the best of our knowledge, this is the first time that the Box-Cox transformation is extended to digital images by exploiting histogram transformation.

## I. INTRODUCTION

It is not uncommon that image-based computer vision algorithms start with a pre-processing phase whereby images are transformed to prepare the data for further processing. Image transformation may embody contrast stretching of intensity values, histogram equalization or its adaptive version, intensity normalization, point-wise operation (e.g., gamma correction), etc. The colours present in an image of a scene supply information about its constituent elements. However, the richness of this information depends very much on the imaging conditions, such as illumination conditions [1] which may significantly degenerate the performance of a variety of computer vision and pattern recognition algorithms.

To eradicate any confusion, we stress -in what follows- that by the term gamma correction, we mean the power-law adjustments performed to improve the quality/contrast of

The associate editor coordinating the review of this manuscript and approving it for publication was Gulistan Raja .

images. Gamma correction, likewise, arcsine transform, are all members of a class of transformations known formally as power transformation which also encompasses the so-called Box-Cox transformation (BCT), a theme that forms the core of this work. The BCT, as a versatile technique, is mostly popular within the statistical and information theory communities. It aims at improving normality of a distribution and equalizing variance (reducing heteroscedasticity) to meet statistical assumptions and improve effect sizes for quantitative analysis of data [2].

Traditionally, BCT is applied to a vector (one-dimensional data) but, to the best of our knowledge, it has not been extended to matrices exhibiting adjacency correlation such as images except in Bicego and Baldo [3], whose work, unfortunately, provides only a cursory overview of the subject. Besides, the generalization to a $d$ -dimensional set of points that they advocate for which typically consists in performing $d$ 1-dimensional transformations, one for each direction of the problem space, is time consuming and not feasible in our case. The other work is that of Lee *et al.* [4] who exploited the parameter lambda ($\lambda$) to further extend the classical mixtures expectation-maximization segmentation to allow generalisation to non-Gaussian intensity distributions for medical MR images.

The rationale behind our approach is not in quest of gaussianity, since images do not always conform to unimodality, but rather to enhance images and boost classes separability (in machine learning context). Of the many techniques currently in vogue for image enhancement, we advocate for the use of our approach both in tandem with machine learning and as a general tool for image enhancement.

### A. MOTIVATIONS

This work is motivated by the scarcity of automatic algorithms that fine-tune the parameter $\lambda$ in gamma power transformation for image enhancement. Power transformations are ubiquitously used in various fields, however, estimating proper values for $\lambda$ remains problematic. For instance, Fattal [5] proposed an algorithm that returns the atmospheric light colour (orientation and magnitude) and stated within the implementation that gamma correction might help orientation recovery where he suggested setting it to 1.5. In Ren *et al.* [6], the authors recommended in their implementation that if the input is very hazy, one can use a large gamma value, but they did not reveal the mechanism. In Berman *et al.* [7]'s implementation, they borrowed gamma values of specific images from Fattal [5]. In Meng *et al.* [8]'s implementation they set $\lambda$ 2 as a regularization parameter.

MATLAB's built-in function, imadjust (ver. 2019a), defaults $\lambda$ to 1 (linear/identity mapping) to dictate the shape of the curve describing relationship of input and output values. MATLAB lets fine-tuning it to the user's arbitrary guess, though the software highlights generic intuitive guidelines to set $\lambda$ without delving into any insights on how to estimate that automatically. Liu *et al.* [9] proposed a dehazing method where $\lambda$ is set to 0.5 in the Gamma correction based pre-processing step. Shi *et al.* [10] realised, as we did, that the traditional power-law transformation has the following disadvantage, increasing $\lambda$ would overcompensate the image's Gamma and thereby darken the processed image while enhancing its contrast. As they did not work out a remedy for such disadvantage, they eventually resorted to using intensity range normalization. These were partially the impetus behind this study.

Therefore, herein, in this article we pinpoint this problem and provide a real-time solution to find the optimal $\lambda$ automatically, which is an essential pre-processing phase in various image processing-based disciplines. This article comes purely to address this issue and propose a solution to it.

### B. CONTRIBUTIONS

In a nutshell, the contributions of this paper can be summarised as follows:

- Indirectly extending the statistical method, BCT, to digital images to establish informed statistical inference on how to estimate image transformation.
- Suggesting a simple yet robust, efficient and inexpensive image enhancement technique that is data dependent (i.e., adaptive) and parameter-free.
- Refining current state-of-the art colour pattern identification algorithm.

The remainder of the paper is apportioned to the following sections: Section II discusses the related work, Section III reviews the BCT algorithm, Section IV discusses the application of BCT to digital images (termed henceforth BCI), Section V brings about the experimental set-up as well as the data sets which are utilized in this study. Section VI is devoted to results and discussion and Section VII concludes this paper.

## II. RELATED WORK

Herein, we list some of the existing and commonly used image enhancement techniques.

### A. CONTRAST LIMITED ADAPTIVE HISTOGRAM EQUALISATION (CLAHE) [11]

In response to the drawback of global histogram equalisation in giving unfavourable results, the CLAHE operation was proposed with two major intensity transformations. The local contrast is estimated and equalized within non-overlapping blocks in the projection image, subsequently, the intensities are normalized at the border regions between blocks through bilinear interpolation. The name contrast-limited refers to the clip limit, which is set to avoid saturating pixels in the image [12].

### B. SUCCESSIVE MEANS QUANTIZATION TRANSFORM (SMQT) [13]

This is an iterative method that can automatically enhance the image contrast. It is capable to perform both a non-linear and a shape preserving stretch of the image histogram.

### C. BRIGHTNESS PRESERVING DYNAMIC FUZZY HISTOGRAM EQUALIZATION (BPDFHE) [14]

This method enhances images by means of calculating fuzzy statistics from image histogram and is built on a prior work.

### D. ADJUSTING IMAGE INTENSITY VALUES (IMADJUST) [15]

In here, we use MATLAB's built-in function which maps the intensity values in a grayscale image to new values. By default, *imadjust* saturates 1% at both top and bottom of all pixel values, resulting in increase of contrast in the output image.

### E. ADAPTIVE GAMMA CORRECTION WITH WEIGHTING DISTRIBUTION (AGCWD) [16]

Huang *et al.*, presented an automatic transformation technique that improves the brightness of dimmed images via gamma correction and probability distribution of luminance pixels.

### F. WEIGHTED VARIATIONAL MODEL (WVM) [17]
This algorithm estimates both the reflectance and the illumination from a given image whereby a new weighted variational model is imposed for better prior representation. The authors claim that their model can preserve the estimated reflectance with more details. However, when we tested it on square matrices of size $2^{11} \times 2^{11}$, it took 76.59 sec to converge on average using the authors' original implementation.

### G. LOW-LIGHT IMAGE ENHANCEMENT (LIME) [18]
The algorithm proposes a simple yet effective low-light image enhancement method where the illumination of each pixel is first estimated individually by finding the max (R, G, B). Subsequently, it refines the initial illumination map by imposing a structure prior on it to produce the final illumination map. Finally, the enhancement is achieved guided by the obtained illumination map.

## III. THE BOX-COX TRANSFORMATION (BCT)
BCT is a parametric non-linear statistical algorithm that is often utilized as a pre-processing channel to convert data to normality, it is credited to Box and Cox [19]. The method is part of the power transform techniques whose quest is to find the parameter lambda, $\lambda$, by which the following log-likelihood is maximized.

$$\mathbf{L}(\lambda) \equiv -\frac{n}{2} \log \left[ \frac{1}{n} \sum_{j=1}^{n} \left( x_j^{\lambda} - \overline{x^{\lambda}} \right)^2 \right] + (\lambda - 1) \sum_{j=1}^{n} \log x_j \quad (1)$$

where $\overline{x^{\lambda}}$ is the sample average of the transformed vector.

There are different attempts to modify this transform, such as those of John and Draper [20] who introduced the so-called modulus transformation and Bickel and Doksum [21] who provided support for unbounded distributions, nevertheless, we prefer to stick to the original definition of the transform as defined in Eq. 2.

$$X(\lambda) = \begin{cases} \dfrac{\chi^{\lambda} - 1}{\lambda}, & \text{if } \lambda = 0, \\ \ln(\chi), & \text{if } \lambda \cong 0. \end{cases} \quad (2)$$

$\forall \chi \in \mathbb{R}_{>0}$, where $\chi$ is a data vector that we wish to transform, and $ln$ is the natural logarithm applied when $\lambda$ approaches zero (i.e., invoked in our case arbitrarily when $\lambda \cong 0.01$). The tested $\lambda$ values are normally in practice bounded (e.g., [-2 2] or [-5 5] are two common ranges).

The BCT's goal is to ensure that the assumptions for linear models are met so that standard analysis of variance techniques may be applied to the data [22]. The algorithm could be a direct possible solution to automatic retrieval of the value of $\lambda$ that somewhat relates to gamma correction. If the parameter $\lambda$ can be properly determined, then each enhanced pixel brightness can be mapped to the desired value and hence contribute to maintaining the overall brightness [23]. The BCT does not change data ordering as per Bicego and Baldo [3].

Obviously not all data can be power-transformed to yield normality, however, Draper and Cox [24] argue that even in cases that no power-transformation could bring the distribution to exactly normal, the usual estimates of $\lambda$ can help regularize the data and eventually lead to a distribution satisfying certain characteristics such as symmetry or homoscedasticity. The latter is especially useful in pattern recognition and machine learning (e.g., Fisher's linear discriminant analysis).

## IV. BOX-COX FOR IMAGES (BCI)
As mentioned earlier, there is a lack of studies that deal with BCT and its power transformation in conjunction with digital images. BCT is an iterative algorithm and applying it to large images would take prohibitively considerable time to converge (e.g., on a square image of size $2^{11}*2^{11}$ it took the BCT algorithm around 10sec to converge on our machine, while operating at the histogram level the time complexity is theoretically size independent, and it took 0.05sec on this image). This feature proves its merit in the big data era where processing large scale image data sets is a concern. The key idea here is to consider the image histogram as a compressed proxy of the entire data matrix since it reflects the estimate of pixel probability distribution as a function of tonal. In this section, we lay down our algorithm in reference to colour images and the application to a grayscale type is encompassed within.

Given a true colour image in the primary red-green-blue (RGB) colour space,

$$\mathbf{F}(u, v) = \{R(u, v), G(u, v), B(u, v)\}, \quad (3)$$

where $(u, v)$ are the pixel spatial coordinates $u \in U$, $v \in V$ and $(U, V)$ are the two image dimensions.

By referring to Eq. 2 and after having an estimate of the parameter $\lambda$ for an input image, we check if the following equality holds:

$$\mathbf{F}(u, v)^{\lambda_{\mathbf{F}}} \cong? \mathbf{F}(u, v)^{\hat{\lambda}_{\chi}},$$

$$\text{where } \chi(i) = \sum_{i=0}^{255} \mathbf{F}_i, \ i \text{ is the gray level, and}$$

$$\mathbf{F} = (0.299R + 0.587G + 0.114B). \quad (4)$$

$\mathbf{F}$ corresponds to the gray level channel as the $YC_bC_r$ colour space calculates it. This colour space is proven to be useful in teasing apart the high frequency signal from the chroma tones that are blended in the RGB space.

We experimentally scrutinized the relationship in Eq. 4. For finding the transformation parameter, *lambda*, whether to derive it from the image matrix, $\lambda_{\mathbf{F}}$, or from the image probability function (a.k.a histogram), $\hat{\lambda}_{\chi}$, (see Eq. 4), we found that the two options yield different gamut enhancement effects in the majority of cases, however, the merits of relying on the histogram are twofold. Our empirical observations indicate the stability as well as the high gain in time complexity when estimating $\lambda$ from the histogram, see Fig 1(c). Fig. 1(a) depicts the Spearman correlation coefficients of both transformations (images are compared) using a sample size of 600 randomly selected natural images acquired by several camera models (correlation between $\hat{\lambda}_\chi$ & $\lambda_\mathbf{F}$ was $r^2 = -0.3022$). Despite the plot seemingly exhibiting an adequate correlation in most cases, the underneath visual impact on the transformed image is not clear from the plot. For example, there are a few instances (e.g., images 64 and 174) when visually examined, Fig. 1(b), they pinpointed the stability of our choice ($\lambda$ histogram). Therefore, we conclude that (w.r.t. Eq.4):

$$\mathbf{F}(u, v)^{\lambda_\mathbf{F}} = \mathbf{F}(u, v)^{\hat{\lambda}_\chi}.$$

Since the BCT may produce values outside of the image permissible dynamic range, therefore, in our case, rescaling the range is invoked which takes the form:

$$BCI = \frac{(\mathbf{F}(u, v) - \min(\mathbf{F}(u, v)))}{\max(\mathbf{F}(u, v))},$$
$$where \ \mathbf{F}(u, v) = \mathbf{F}(u, v)^{\hat{\lambda}_\chi} \quad (5)$$

## V. EXPERIMENTAL SET-UP

The extension of the Box-Cox transformation to digital images would not be complete without exploring how the estimation of $\lambda$ affects some image-domain specific applications. This section shall provide insight into two dominant areas: image enhancement and image colour pattern classification using a recent pre-trained model. In the below experiments, compressed images (i.e., JPEG, JPG), are converted to lossless compressed type (.png) before carrying out any analysis to prevent re/compression artifacts contaminating the statistical conclusions.

### A. IMAGE ENHANCEMENT

Testing for the capability of our proposed approach, BCI, against commonly used methods is carried out, for this task, using the *Phos II* data set along with images collected from the illumination dataset [25]. *Phos II* [26] is a colour image database of 15 scenes captured under different illumination conditions. More concretely, every scene of the dataset contains 15 different images: 9 images captured under various strengths of uniform illumination, and 6 images under different degrees of non-uniform illumination. The images contain objects of different shape, colour and texture.

#### 1) TESTS AND EVALUATION METRICS

Probability Distribution Test on Simulated Data: As a sanity test, we first create a synthetic image (a gradient map) where each row is a vector that defines 257 equally spaced points between 0 and 1, see Fig. 2.

We then contrast our proposed approach, BCI, to enhancements using the methods reported in section II. To assess the goodness of fit, the QQplot (quantile-quantile plot) is utilised which plots the quantiles of the input vector data against the theoretical quantiles of the distribution specified by pd (probability distribution). If the empirical distribution conforms to pd, then data points shall fall on a straight line. Our choice of pd landed on the Rayleigh distribution for the very reason that it is commonly used in imaging technology [27]–[30].

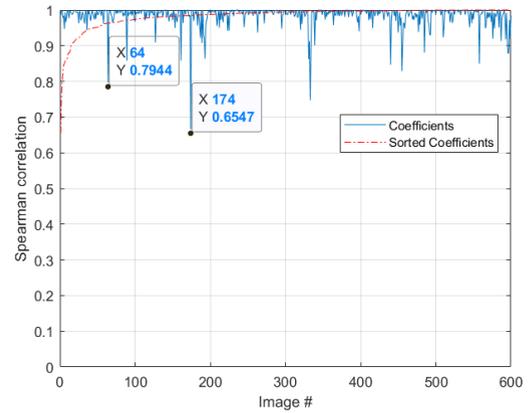

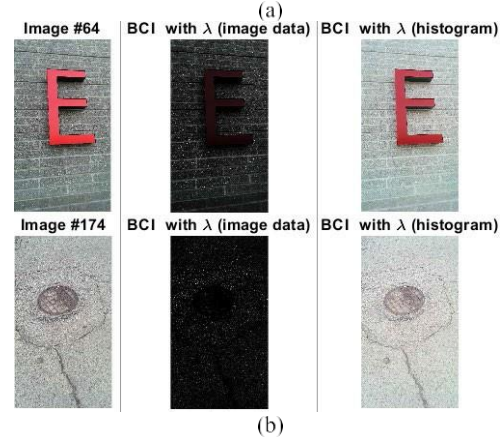

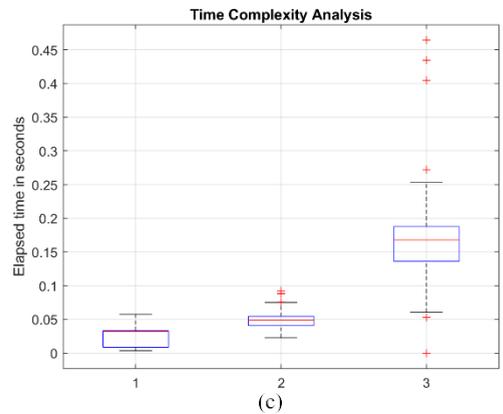

FIGURE 1. Time complexity and correlation coefficients between 600 transformed images using $\hat{\lambda}_\chi$ and $\lambda_\mathbf{F}$. (a) the plot depicts the correlation between BCI enhanced images with $\lambda$ derived from the image histogram and their counterparts with $\lambda$ derived from the image intensity values. (b) visualisation of the transforms on images having the lowest correlation values (i.e., #64 and #174 in a). (c) time complexity of estimating lambda from the histogram (label 1) as compared to estimating it from image data resized to 64xNaN (label 2) and 256xNaN (label 3).

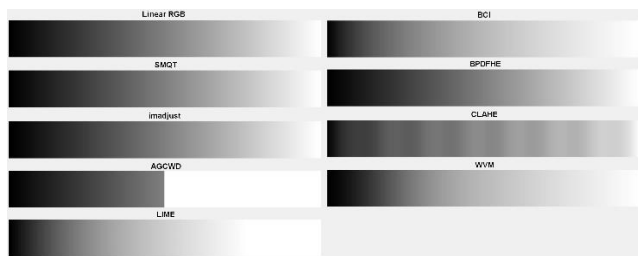

**FIGURE 2.** The image corrected gamut turns brighter more swiftly in (BCI) than the linear colour space and the other methods.

The *Rayleigh* distribution is a special case of the *Weibull* distribution and its probability density function is formally defined as:

$$\rho\,(x|\delta) = \frac{x}{\delta^2} e^{\frac{-x^2}{2\delta^2}} \tag{6}$$

where $\delta = \sqrt{\frac{1}{2n} \sum_{i=1}^{n} x_i^2}$ is a scale parameter of the distribution.

**QQ Plot of Sample Data versus Rayleigh Distribution**

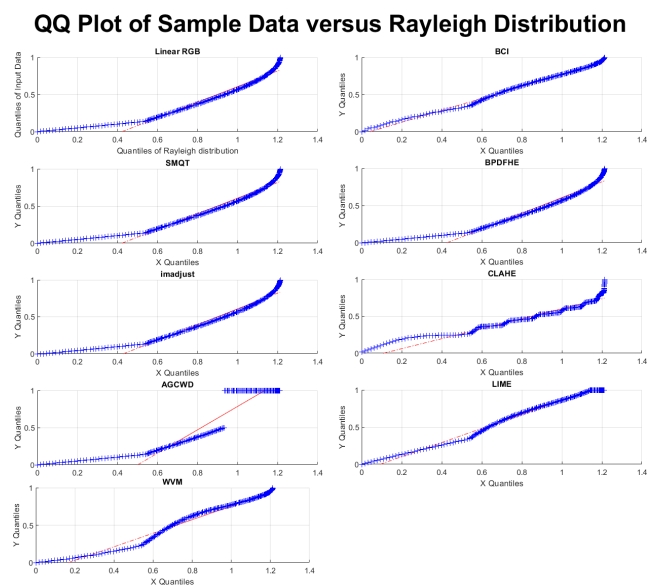

**FIGURE 3.** Rayleigh distribution fit test using QQplots on the synthetic images shown in Fig. 2. Additional tests can be found on our webpage (see the link in Section VI).

In Fig. 3, QQplots of the synthetic image and the enhancements using the eight methods (as shown in Fig. 2). The overall impression one gets from this visualisation of assessing goodness of fit is that BCI, as compared to other methods, is the output that fitted most on the probabilistic line.

In the above experiment, we noticed that the AGCWD's output was far from what we expected from this algorithm. This observation triggered us to extend our experiments by varying the vector length to observe the algorithm's behaviour. The AGCWD re-affirmed our observation, see Section VI and the web-link therein.

## 2) QUANTITATIVE EVALUATION METRICS
In this sub-section, we highlight the different statistical metrics that we utilise. The intention here is not to go into details as these metrics are well established popular measurements. Quantitative evaluation of contrast enhancement is not an easy task. Huang *et al.* [16] attributed that to the absence of an acceptable criterion by which to quantify the improved perception, quoting also [31], [32]. However, since then, a couple of image quality evaluator metrics have been proposed and are currently widely used. Hence, to gauge image enhancement efficiency, the so-called blind image quality metrics are adopted.

### a: NATURALNESS IMAGE QUALITY EVALUATOR (NIQE) [33]
This metric compares a given image to a default model derived from natural scene statistics. The score is inversely correlated to the perceptual quality, in other words, a lower score indicates better perceptual image quality.

### b: PERCEPTION BASED IMAGE QUALITY EVALUATOR (PIQE) [34]
This metric calculated the score through block-wise distortion estimation. The score is inversely correlated to the perceptual quality.

### c: BLIND/REFERENCELESS IMAGE SPATIAL QUALITY EVALUATOR (BRISQUE) [35]
This metric compares a given image to a support vector regressor model trained on a set of quality-aware features and corresponding human opinion scores. The score is inversely correlated to the perceptual quality.

## VI. RESULTS AND DISCUSSIONS
Herein, we warrant the merits of the proposed approach (BCI) by conducting quantitative comparisons. The results give us a cue that BCI can be a potential alternative for existing methods. BCI time complexity should not be a concern since the algorithm, as we stated earlier, operates on image histogram ($<= 256$ points to process) to derive $\lambda$. BCI, like any other image enhancement algorithm, alters colour gamut. Therefore, for studies that are interested in the relationship between colours (assuming quantitatively accurate intensity values), such as the case in studies on $\beta$-cells promotion of insulin secretion or protein expression levels [36], should keep this fact in mind when dealing with image enhancement in general. The numerical output that we report here go into three directions, first the image enhancement domain, second, the vivid research area of image de-hazing [37], and finally the area of machine learning for image classification (deep/transfer learning for image classification).

### A. IMAGE ENHANCEMENT
In this section, we demonstrate the integrity and stability of our approach against two tests, namely, quality enhancement test and colour pattern segmentation test. In the first experiment, we selected 550 images exhibiting non-uniform lighting and contrast conditions. Images are of different sizes and are stored in RGB format. Table 1 tabulates the obtained results averaged across the entire set. It is evident that, on average, BCI outperforms all methods in quality assessments (i.e., NIQE, PIQE, BRISQUE). It is important to know that a couple of the methods shown in Table 1 operate only on single channel images (e.g., CLAHE), consequently, we convert the input image to HSV where these algorithms operate on the V channel then the image is reverted back to RGB space. It is interesting to see, from this analysis, that the algorithm SMQT retains image statistics which results in it having the same scores as the original image. Twelve randomly selected samples drawn from the 550 set are shown in Fig. 4.

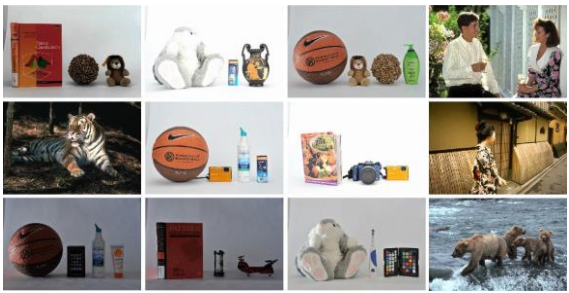

**FIGURE 4.** Randomly selected samples from the 550-test set used to generate Table 1.

Given both extremes, BCI can be singled out for giving consistent favourable results in both cases. As for the SMQT, imadjust and BPDFHE, the contrary is true, they are prone to severe performance degradation under low exposure. This observation is consistent across additional experiments we conducted as a sanity check. Ultimately, this may be the best use of BCI transformation technique for those cases when inferences on the optimal transformation can be affected by exposure uncertainty.

**TABLE 1.** Image enhancement performance evaluation.

| Method \ Metric | NIQE | PIQE | BRISQUE |
|---|---|---|---|
| Original (RGB) | 3.3500 | 38.0827 | 27.5242 |
| SMQT [13] | 3.3500 | 38.0827 | 27.5242 |
| Imadjust [15] | 3.3282 | 38.9011 | 25.2030 |
| CLAHE [11] | 3.4870 | **37.4277** | 24.7865 |
| BPDFHE [14] | 3.4718 | 41.7140 | *23.9606* |
| AGCWD [16] | 3.3341 | 38.8288 | 24.7123 |
| LIME [18] | 3.6035 | 40.8890 | 27.4957 |
| WVM* [17] | *3.2113* | 38.3200 | 25.7272 |
| BCI (Proposed) | **3.1866** | *37.8784* | **22.9973** |

(*) The WVM algorithm is very time consuming -impractical-, therefore, we shall henceforth cease using it in the experiments.

Results on a grayscale image are shown in Fig. 5. To not clutter this paper with images, higher resolution visual qualitative comparisons on RGB still images and on simulated synthetic data (animation) that define different vector lengths

(see Sec IV) are all furnished online through the following page: http://www.abbascheddad.net/BCI.html.

It is observed that the LIME algorithm (Fig. 5j) malfunctions around bright light regions in the image (i.e., exaggerates the oversaturated/bright areas), this phenomenon was also observed on additional tests that we conducted (data not shown). Another property of BCI is its ability to make data distribution less asymmetric as compared to existing methods. We tested this property on 600 randomly selected natural images by using descriptive statistics, the skewness (*Skew*) and kurtosis (*Kurt*). The average results are depicted in Table 2.

**TABLE 2.** Skew and kurtosis tests.

| | Skew | Skew* | Kurt | Kurt* |
|---|---|---|---|---|
| Original | 3.3599 | 3.3402 | 34.8535 | 34.2114 |
| SMQT [13] | 3.3599 | 3.3402 | 34.8535 | 34.2114 |
| Imadjust [15] | *3.2778* | *3.2586* | *33.9541* | *33.3295* |
| CLAHE [11] | 3.4555 | 3.4352 | 45.7557 | 44.9018 |
| BPDFHE[14] | 3.8149 | 3.7925 | 43.7525 | 42.9376 |
| AGCWD[16] | 5.2511 | 5.2203 | 59.8653 | 58.7374 |
| LIME [18] | 13.8999 | 13.8183 | 215.7154 | 211.5601 |
| BCI | **2.8020** | **2.7855** | **21.7072** | **21.3204** |

(*) after adjusting for the systematic bias (based on the sample size)

### B. COLOUR PATTERN SEGMENTATION

Pixel-wise colour pattern segmentation has been a long-standing research problem. Weijer *et al.* [38], [39], proposed a handy algorithm where colours are learned from real-world noisy data. To avoid manual labelling, their learning model is trained on colour images retrieved from Google image search engine. The algorithm can recognise colour patterns belonging to 11 colour gamut, namely, black, blue, brown, grey, green, orange, pink, purple, red, white and yellow. In this experiment, we show that BCI does improve the performance of Weijer *et al.*'s method if incorporated prior to segmentation. In Fig. 6, we provide three examples, showing challenging synthetic chromaticity images.

### C. OTHER APPLICATIONS

This section delves into some contemporary fields that can take advantage of the developed BCI method. Namely, we will examine a face recognition problem using deep learning and another vivid computer vision area known as image dehazing (removing haze from captured images).

#### 1) DEEP LEARNING (FACE RECOGNITION)

In these experiments we report the average mean of running 10-fold cross-validation (70% training, 30% test).

(*The extended Yale Face Database B*): This set contains 1922 images of 38 human subjects under 9 poses and 64 illumination conditions [40]. The images are of dimensions 168 $\times$ 192. The variation in illuminance in this data set forms an ideal platform to test the BCI's performance. In here we use the renown pretrained deep learning model, AlexNet, for what is termed as transfer learning. We trained a Support

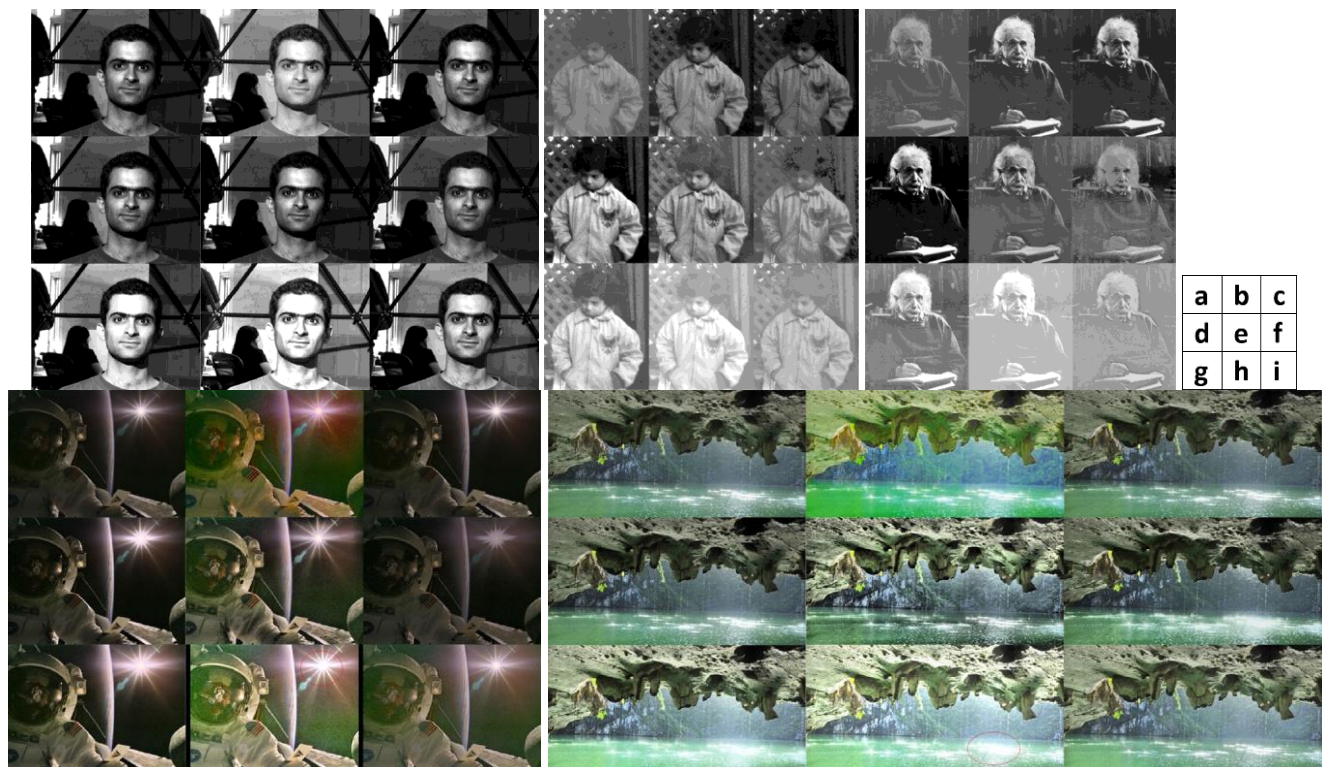

**FIGURE 5.** Enhancement of grayscale images. (a) Original low-contrast image. Result of using (b) BCI, (c) SMQT [13], (d) imadjust [15], (e) CLAHE [11], (f) BPDFHE [14], (g) AGCWD [16], (h) LIME [18] red circles added to highlight exaggerated bright areas -astronaut and lake-, and (i) WVM [17].

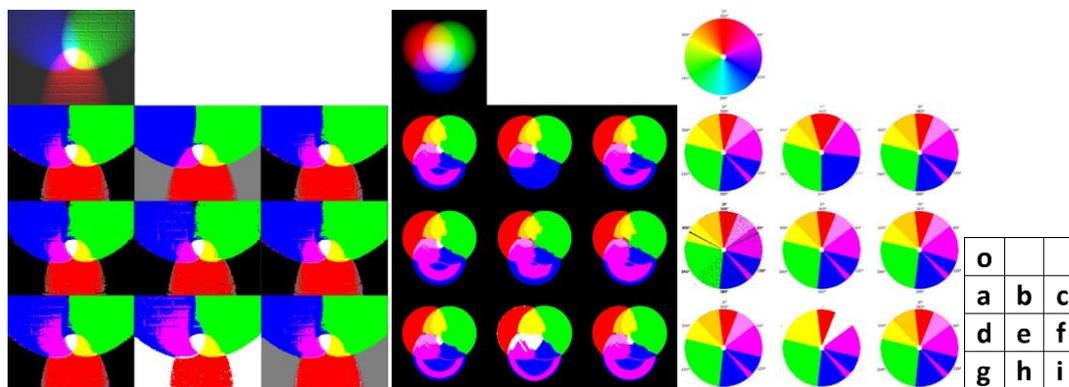

**FIGURE 6.** Three examples of pixelwise colour pattern segmentation[1]. (o) Segmentation-challenging synthetic image. (a) Results obtained with the native code of [38], results obtained with: (b) [38] +BCI, (c) [38] +SQWT [13], (d) [38] +imadjust [15], (e) [38] +CLAHE [11], (f) [38] +BPDFHE [14], (g) [38] +AGCWD [16], (h) [38] +LIME [18], (i) [38] +WVM [17].

Vector Machines model (SVM) as a classifier for the Yale Database with and without image enhancements. The dataset was divided into 70%, 30% for training and test, respectively. To eliminate any overfitting and/or biases in the selected samples, random selection was deployed, and the process was repeated 10 times. We then reported in Table 3, the average accuracy and the best AUC (the area under the ROC Curve) of each method.

### 2) IMAGE DEHAZING

Visual quality can be decreased substantially due to adverse weather conditions (e.g., fog), man-made air pollution (fire fume, smoke-bombs by football fans, lachrymator when combating riots), etc. The optical field of science that deals with restoration of degraded photographs captured during such situations, is known as image de-hazing. It is a vivid research area, as evidenced by its presence in one of the major conferences on computer vision and pattern recognition, namely, CVPR'2019 workshop on Vision for All Seasons: Bad Weather and Night-time (https://vision4allseasons.net/).

The additional statistical metrics that we utilised in the experiments reported in this sub-section for the de-hazing scenario are the reference-based image quality metrics. The Peak Signal-to-Noise Ratio (*PSNR*), the Structural Similarity Index (*SSIM*), the Information Content Weighted PSNR (*IWPSNR*) [41], the Information Content Weighted SSIM

---

[1]Full resolution available online: http://www.abbascheddad.net/BCI.html

(*IWSSIM* ) [41], and the Pearson Correlation Coefficient (*Corr*).

We selected the top two methods (i.e., LIME and BCI) from Table 3. Additionally, we run a comparison of six existing image de-hazing methods (data not shown), namely, Ren *et al.* [6], Berman *et al.* [7], Galdran [42], Fattal [43], Meng *et al.* [8] and He *et al.* [44], we found that Galdran [42] outperformed other methods based on PSNR and Corr values. Additionally, Galdran [42] is the most recent method published in 2018. Therefore, we selected it to test the added value that LIME and BCI may introduce.

The advantage of our approach in boosting the performance of image de-hazing algorithms on the O-Haze dataset [45] is depicted in the results shown in Table 4. Although, the improvement is consistent across all metric, it is a mild improvement (except for the correlation value). We also noticed that if the dehazing method in [9] is pre-processing using LIME, the latter would introduce a reversed effect. This again, advocated for the stability and utility of the proposed BCI approach.

**TABLE 3.** Results of image classification on the yale face database B.

| AlexNet TransL SVM | | |
|---|---|---|
| **Method \ Metric** | **Accuracy (mean)** | **AUC** |
| Baseline | 0.8711 ±0.0094 | 0.9401 |
| SMQT [13] | 0.8734 ±0.0108 | 0.9446 |
| Imadjust[15] | 0.8782 ±0.0090 | 0.9425 |
| CLAHE [11] | 0.9142 ±0.0124 | 0.9650 |
| BPDFHE [14] | 0.8778 ±0.0079 | 0.9420 |
| AGCWD [16] | 0.8945 ±0.0111 | 0.9527 |
| LIME [18] | 0.9308 ±0.0117 | 0.9720 |
| WVM [17] | 0.9005 ±0.0126 | 0.9579 |
| BCI (Proposed) | **0.9496 ±0.0064** | **0.9768** |

**TABLE 4.** Quantitative evaluation of all the 45 set of images of the O-Haze dataset. This table presents the average values of the seven quality metrics, over the entire dataset.

| Metric\ Method | Baseline [9] | BCI+Baseline [9] | LIME+Baseline [9] |
|---|---|---|---|
| PSNR | 16.6655 ±2.1453 | **16.7408** ±3.0026 | 11.5025 ±1.9707 |
| IWPSNR | 63.9008 ±2.6778 | **65.3192** ±2.3188 | 62.1111 ±2.3551 |
| SSIM | 0.7074 ±0.1030 | **0.7316** ±0.0910 | 0.6395 ±0.0831 |
| IWSSIM | **0.9914** ±0.0056 | 0.9906 ±0.0079 | 0.9789 ±0.0098 |
| Corr | 0.6411 ±0.2659 | **0.7210** ±0.2115 | 0.6223 ±0.2479 |

(*) Row-wise, the highest score of each metric is given in bold (mean ± standard deviation).

## VII. CONCLUSION

In this paper, we propose a new approach to enhance images by extending the renowned Box-Cox transformation to 2D data. Since Box-Cox algorithm stems from statistical and probability theories and since it is conducted to, among other benefits, stabilize the variance in one dimensional data (e.g., a vector of covariate/confounding variables), extra vigilance should be taken when tackling digital images. Our approach, termed herein BCI, precludes the need to arbitrarily estimate the parameter $\lambda$ in Gamma correction or the need to find limits to contrast stretch an image. When this approach was conceived, we tried to not involve regularization parameter controls into our algorithm to reduce complexity and ease replication of results. The proposed scheme is simple and fast, does not require any model training, and we believe that it can complement other existing image enhancement algorithms.

The results land credibility to the efficiency of our proposed approach and show its stability and robustness compared to commonly used contrast enhancement techniques. Subsequently, we support our approach by improving the performance of the state-of-the art colour learning algorithm and a deep learning algorithm (see Section VI). This paper warrants a succinct description of the proposed approach, however, due to the page limit we have omitted other promising results in other domains which could have otherwise instilled credibility even more in the notion of BCI.

The Box-Cox algorithm as a well-adopted statistical and probabilistic method, is shown in this study to retain its fidelity even on two-dimensional data (i.e., digital images). One of the aims of this paper is to rekindle interest in the Box-Cox algorithm in conjunction with image enhancement. In a wider context, this optimisation algorithm might even help leverage the results of other enhancement algorithms that depend on the parameter $\lambda$, such as [10], and/or those setting it arbitrarily for gamma correction [5]–[9], and in other areas which we did not cover here such as image retrieval where informative features are sought [46]. There are some attempts to devise new methodologies to estimate $\lambda$ for 1-dimensional data transformation, like the work of [47], however, this proposal comes to create an accrual of evidence regarding the utility of the renowned Box-Cox transformation in the imaging field. A possible road map for future work could be to examine the performance of the proposed BCI on different colour space transformations, or in other domain specific applications (e.g., integration into deep learning architectures, image fusion, etc.).

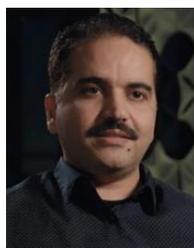

**ABBAS CHEDDAD** (Senior Member, IEEE) received the Ph.D. degree (Hons.) from the University of Ulster, U.K., in 2010. He has held research positions at several universities in Sweden, such as, Umeå University and Karolinska Institute, where he focused his research on medical image analysis (disease risk prediction and optical projection tomography). He is leading a research group on big data analytics for image processing. Currently, the group is collaborating, research-wise, with three companies, SONY Mobile Communications AB, Lund, Arkiv Digital AB, Mariestad, Sweden, and GKN Aerospace AB (the world's leading multi-technology tier 1 aerospace supplier) by addressing practical industrial problems. He is an Associate Professor (Docent) with the Blekinge Institute of Technology (BTH), Sweden. He has in records, one book, one book chapter (invited), two granted patents, and more than 60 journal articles and conference papers. He is a member of the IEEE Signal Processing Society and an ACM Distinguished Speaker. He received several awards, including the 25K Award for New Entrepreneurs in the hi-tech category sponsored by Northern Ireland Science Park (NISP) and has acquired grants that total up to 384 000 €. He was the Chair of three international conferences/workshops, a PC member in dozens of conferences, and has been invited for talks at several venues.

• • •